\documentclass[aps,floats,twocolumn]{revtex4}
\input epsf
\usepackage{amsmath,amssymb}
\usepackage{graphicx}

\newcommand{\be}{\begin{equation}}
\newcommand{\ee}{\end{equation}}

\begin{document}

\title{Constraints on a New Post-General Relativity Cosmological Parameter}

\author{Robert Caldwell}
\affiliation{ Department of Physics and Astronomy, Dartmouth College, 6127 Wilder 
Laboratory, Hanover, NH 03755 USA}

\author{Asantha Cooray}
\affiliation{Center for Cosmology, Department of Physics and Astronomy, 
  University of California, Irvine, CA 92697 USA}
  
\author{Alessandro Melchiorri}
\affiliation{Physics Department and Sezione INFN, University of  Rome ``La Sapienza'',
P.le Aldo Moro 2, 00185 Rome, Italy}
 
\date{\today}

\begin{abstract} 

A new cosmological variable is introduced which characterizes the degree of departure from
Einstein's General Relativity (GR) with a cosmological constant. The new parameter, $\varpi$,
is the cosmological analog of $\gamma$, the parametrized post-Newtonian variable which
measures the amount of spacetime curvature per unit mass. In the cosmological context,
$\varpi$ measures the difference between the Newtonian and longitudinal potentials in response
to the same matter sources, as occurs in certain scalar-tensor theories of gravity.
Equivalently, $\varpi$ measures the scalar shear fluctuation in a dark energy component. In
the context of a ``vanilla" LCDM background cosmology, a non-zero $\varpi$ signals a departure
from GR or a fluctuating cosmological constant. Using a phenomenological model for the time
evolution $\varpi=\varpi_0 \rho_{\rm DE}/\rho_{\rm M}$ which depends on the ratio of energy
density in the cosmological constant to the matter density at each epoch, it is shown that the
observed cosmic microwave background (CMB) temperature anisotropies limit the overall
normalization constant to be $-0.4 < \varpi_0 < 0.1$ at the 95\% confidence level. Existing
measurements of the cross-correlations of the CMB with large-scale structure further limit
$\varpi_0 > -0.2$ at the $95 \%$~CL.  In the future, integrated Sachs-Wolfe and weak lensing
measurements can more tightly constrain $\varpi_0$, providing a valuable clue to the nature of
dark energy and the validity of GR.  

\end{abstract}

\maketitle

\section{Introduction}
 
There is a gap in our understanding of the cosmos: we do not understand the cause of the
accelerated cosmic expansion and the source of most of the energy of the universe. In the
context of Einstein's General Relativity (GR), the evidence suggests the existence of some
form of dark energy, such as a cosmological constant or quintessence. (For a review see
Refs.~\cite{Peebles:2002gy,Padmanabhan:2002ji,Sahni:1999gb} and references therein.)
Alternatively, novel gravitational phenomena on the largest, cosmological scales may be
responsible for the observations ({\it e.g.}
Refs.~\cite{Carroll:2003wy,Carroll:2004de,Ishak:2005zs,Lue:2003ky,Zhang:2005vt,Knox:2006fh,Uzan:2006mf,Huterer:2006mv}).
One of the outstanding challenges to physics and astronomy is to discover the answer to this
problem. Our goal is to reformulate the degeneracy between dark energy and novel gravitational
phenomena into a simple formalism. In doing so, we aim to extend the parametrized
post-Newtonian formalism (see also
Refs.~\cite{Tegmark:2001zc,Sealfon:2004gz,Shirata:2005yr,Jaekel:2005qe,Jaekel:2005qz}), which
has served so well to test GR within the solar system, to cosmological scales.

The plan of the paper is as follows. In section II we motivate and introduce the new
cosmological parameter, $\varpi$, making a connection with the solar system PPN parameters. We
show that this parameter may characterize new gravitational physics, but is also degenerate
with shear perturbations which we attribute to the dark energy. In section III we propose a
model for the parameter, $\varpi$, detail our procedure for the computation of cosmological
observables, and present a sample of the consequences of $\varpi \neq 0$. Our procedure is not
unique but has practical advantages. In sections IV and V we present constraints on $\varpi$
and forecasts using the cosmic microwave background (CMB), cross-correlations with large-scale
structure, and weak lensing. A summary with conclusions is given in section VI. We include an
appendix explaining how dark energy with $w=-1$ can support fluctuations.

\section{Cosmic PPN} 

General Relativity (GR) tells us precisely how mass and energy curve spacetime. In other
theories of gravity, the relationship between between matter and curvature is different.  By
presuming GR and looking for contradictions we can hope to test the consistency of GR.  
Post-Newtonian variables are used to quantify the behavior of gravity and departures from
General Relativity. (See Refs.~\cite{WillBook,Will:2001mx} for a thorough review.) The
Eddington-Robertson-Schiff metric 
\begin{eqnarray}
ds^2 &=& -\left[1 - 2 \alpha \frac{G M}{r} + 2(\beta - \alpha \gamma) 
\left(\frac{G M}{r}\right)^2 + ... \right] dt^2 \cr
&+& 
\left[ 1 + 2 \gamma \frac{G M}{r} + ...\right] dr^2 + r^2 d\Omega^2
\label{eqn:PPNmetric}
\end{eqnarray}
can be used to parametrize the way a point mass, $M$, curves space.  The Newtonian potential
sets Newton's constant so that it is convention to set $\alpha = 1$ at the present day. A more
detailed metric and system of parameters is possible,  allowing for the motion of the source,
though is unnecessary at this stage. The parameter $\beta$ measures the amount of
non-linearity in gravitational superposition, and $\gamma$ measures how much curvature is
produced per unit rest mass. In GR, $\beta = \gamma = 1$. The most recent limits are $\beta -1
= 1.2\, (\pm 1.1) \times 10^{-4}$  from lunar laser ranging \cite{Williams:2004}, and 
$\gamma-1 = 2.1\, (\pm 2.3)\times 10^{-5}$ measured from the time-delay of signals from the
Cassini spacecraft \cite{Bertotti:2003}.  Einstein's General Relativity is in excellent
agreement with observations within the solar system. 

These measurements place very tight constraints on extensions of GR. However, it is possible
to imagine that the underlying theory of gravitation resembles GR on solar system scales, but
departs on larger, cosmological scales. The post-Newtonian variable $\gamma$ is very close to
$1$ within the solar system, but may depart on larger scales or vary with cosmic age.  With
the notion in mind that dark energy phenomena may be a result of a departure from GR, it is
tempting to speculate that a departure from GR hinges on the local ratio of dark energy to the
matter density.  Using standard cosmological numbers, this ratio is
\begin{equation}
\langle\rho_{DE}\rangle / \langle \rho_{M}\rangle =  \Omega_{DE}\frac{3 H^2_0}{8\pi G}/
(0.1 M_\odot / {\rm pc}^3) \approx 10^{-6}.
\label{eqn:localratio}
\end{equation}
Consequently, one may speculate that an improvement in solar system tests by an order of
magnitude may reveal faults in GR. Alternatively, it may be easier to search for such faults
using cosmological phenomena, such as the CMB, the growth of large-scale structure
\cite{Knox:2006fh,Huterer:2006mv}, and gravitational lensing
\cite{Bekenstein:1993fs,White:2001kt,Bolton:2006yz}.

To adapt the post-Newtonian formalism to cosmology, we replace the Eddington-Robertson-Schiff
metric with the perturbed FRW metric
\begin{equation}
ds^2 = a^2(\tau)\left[ -(1 + 2 \psi)d\tau^2 + (1 - 2 \phi) d\vec x^2\right].
\end{equation}
Hereafter we employ the notation and conventions of Ma \& Bertschinger (MB) \cite{Ma:1995ey}.
It is convenient to work in the conformal-Newtonian gauge if we wish to maintain the simple
connection to the Newtonian potential as appears in the Poisson equation. If the gravitating
source is a point mass, then we mean to identify $\psi = -\alpha G M/r$ and $\phi = -\gamma G
M/r$. so that measurements of the Newtonian and longitudinal gravitational potentials can be
interpreted as a test of the strength of gravity ($\alpha$) and the amount of curvature
produced per unit rest mass ($\gamma$). Hence, we introduce a new parameter, $\varpi$,
implicitly defined by the equation 
\begin{equation}
\psi = (1 + \varpi)\phi,
\end{equation}
which describes the departure from GR in the cosmological context. In terms of PPN parameters,
$\varpi = \frac{\alpha}{\gamma} - 1$ which translates to $\varpi \approx 1-\gamma$ for a weak
departure from GR. Reexpressing the results of Ref.~\cite{Bolton:2006yz} on the comparison
between lensing and dynamical measurements of galaxy masses, then  $\varpi = 0.02 \pm 0.07$
(68\% CL) on tens of kiloparsec scales. The same analysis can be extended to  a few hundred
kiloparsec scales by making use of both strong and weak lensing measurements towards galaxy
clusters combined with dynamical mass measurements of same clusters. Using the comparison
between weak lensing and X-ray based masses in Ref.~\cite{Wu98}, we find  $\varpi = 0.03 \pm
0.10$ (68\% CL). At megaparsec scales corresponding to cosmological observations,
unfortunately, there are no useful limits placed on $\varpi$.

A difference in the two gravitational potentials can also occur in GR when the gravitating
source has anisotropic stress or shear. Specifically, the off-diagonal components of the
perturbed Einstein equation yield
\begin{equation}
k^2(\phi - \psi) = 12 \pi G a^2 (\rho + p)\sigma,
\end{equation}
where $\rho,\,p$ are the energy density and pressure of the fluid component giving rise to the
shear. Of course, there are other sources of shear, due to the contribution of relativistic
species such as photons or neutrinos. Including the effects due to a departure from GR we
obtain 
\begin{equation}
k^2(\phi - \psi) = 12 \pi G a^2 (\rho + p)\sigma|_{\gamma,\nu} - \varpi \, k^2  \phi.	
\end{equation}
Note that the difference between the two potentials, as well as individual shear components,
are gauge-invariant. In the synchronous gauge, the potentials are
\begin{equation}
\phi = \eta - {\cal H} \alpha,\quad \psi = \dot\alpha + {\cal H}\alpha,\quad
\alpha = \frac{1}{2 k^2}(\dot h + 6 \dot\eta)
\end{equation}
where the dot indicates a derivative with respect to conformal time. Consequently, the shear
parametrization equation may be recast in terms of the evolution equation for $\alpha$:
\begin{equation}
\dot\alpha + (2 + \varpi){\cal H}\alpha - (1 + \varpi)\eta =
-12 \pi G a^2 (\rho + p)\sigma|_{\gamma,\nu}.
\label{eqn:alphadot}
\end{equation}
Hence, the static PPN-like relation between potentials in the conformal-Newtonial gauge
becomes dynamical in the synchronous gauge. In the concordance cosmological model, which
includes Einstein's cosmological constant within GR, photon and neutrino shear are negligible
in the late universe. Hence, an observed difference between the potentials signals some
non-standard behavior.

This behavior may be seen in gravitational phenomena such as the deflection of light. The
deflection angle experienced by a beam of light  moving through a potential $\phi$ now takes
the form $\delta\theta = -2(1+\varpi)\phi$. In the cosmological setting, we can find the
lensing potential for the deflection of photons by adapting the results of Acquaviva {\it et
al} \cite{Acquaviva:2004fv}, which amounts to replacing their $\Xi$ with $\varpi\phi$, to
obtain
\begin{equation}
\varphi (\hat n) = \int_0^{\chi_\infty} d\chi \, g'(\chi) 
\left[ (2 + \varpi) \phi(\hat n,\chi)\right].
\label{eqn:lens}
\end{equation}
Here $\chi$ is the comoving distance from the observer, $g'(\chi) =
\chi\int_\chi^{\chi_\infty} d\chi' (1 - \chi/\chi') n(\chi')$, and $n(\chi)$ gives the 
distribution of background sources. This means the pattern of lensing of background galaxies,
CMB, or any field of photons, will be influenced by $\varpi \neq 0$. 

Similarly, the integrated Sachs-Wolfe effect (ISW) is affected by
$\varpi$:
\begin{equation}
\frac{\delta T}{T}(\hat n)|_{ISW} = \int d\chi\,  
\left((2 + \varpi)\phi\right)_{,\tau}(\hat n,\chi).
\label{eqn:isw}
\end{equation} 
Hence, combined measurements of the CMB with probes of large-scale structure, which are
sensitive to $\phi$ through the Poisson equation, can be used to isolate the effects of
$\varpi$.  

Our focus on the {\it gravitational slip} between the potentials is motivated in part by the
recent work of Bertschinger \cite{Bertschinger:2006aw}, as well as numerous investigations of
scalar-tensor and generalized gravitational theories ({\it e.g.}
\cite{Chen:1999qh,Esposito-Farese:2000ij,Riazuelo:2001mg,Nagata:2002tm,Nagata:2003qn,Acquaviva:2004fv,Schimd:2004nq}) and
dark energy with anisotropic stress \cite{Koivisto:2005mm,Battye:2006mb}. Our intention is to
use these existing tools and techniques, not to explore exotic theories and models, but to
examine simple departures from GR with a cosmological constant.

\section{Cosmic Model}
 
It is evident that $\varpi$ can be used to parametrize departures from Einstein's GR.  A
specific prediction for the gravitational slip in the cosmological context is obtained for a
given theory. For example, the difference between the two conformal-Newtonian gauge potentials
is $\phi - \psi = \delta(\ln L_{,R})$ in a theory of gravity described by a generalized
Lagrangian $L(R,\varphi)$ \cite{Acquaviva:2004fv}, where $R$ is the Ricci scalar curvature and
$\varphi$ is the gravity-coupled scalar field.  The relation between a fourth-order
gravitational theory and PPN parameters $\gamma,\,\beta$ is given in
Ref.~\cite{Capozziello:2006fa}. The gravitational slip in a scalar-tensor theory including a
scalar coupling to the Gauss-Bonnet Lagrangian is analyzed in Ref.~\cite{Amendola:2005cr}. In
the Dvali-Gabadaze-Porrati (DGP) brane-world model of gravity \cite{Dvali:2000rv,Lue:2005ya},
recent studies of structure formation \cite{Koyama:2005kd} derive a gravitational slip $\phi -
\psi =  8\pi G a^2 \rho (\delta_m + 3 {\cal H V}_m/{k^2}) / \beta k^2$. Here $\beta$ is a
dimensionless coefficient which relates the five-dimensional cross-over scale, $r_c$, to the
present-day Hubble constant, whereby $\varpi \approx {\cal O}(1/\beta)$. More examples can
certainly be found. We would like to find a phenomenological description which captures the
basic effects. The advantage we see is that we can streamline the search for non-GR behavior;
if we find evidence for $\varpi \neq 0$ then we can focus on particular theories.

In general, we expect dark energy effects and the corresponding gravitational slip to turn on
at late times. Hence, we propose a simple extension of the concordance, $\Lambda$CDM
cosmological model. 1) We assume that the dark energy phenomena resembles a cosmological
constant such that the background evolution of the expansion scale factor is driven by a
component with equation-of-state $w=-1$. This maintains consistency with all observational
constraints based on classical distances. 2) We assume an evolution for the gravitational slip
which grows large as the role of dark energy grows:
\begin{equation}
\varpi = \varpi_0 \frac{\rho_{DE}}{\rho_{M}} = \varpi_0
\frac{\Omega_{DE}}{\Omega_{M}}(1+z)^{-3}.
\label{eqn:varpi}
\end{equation}
Our immediate goal is to constrain $\varpi_0$. Based on our estimate of the local ratio
between dark energy and matter (\ref{eqn:localratio}), we expect $|\varpi_0| \lesssim 10$ in
order to be consistent with the solar system limits on $\gamma$ \cite{Bertotti:2003}.  Next,
using the approximation that ${\rho_{DE}}/{\rho_{M}} \sim 10^{-4}$ on the scale of a galaxy
halo, then the limits of Refs.~\cite{Bolton:2006yz,Wu98} give $|\varpi_0| \lesssim 100$. The
cosmological ratio, however, only grows significant at late times as the inferred dark energy
comes to dominate. The source of $\varpi$, the slip between $\phi$ and $\psi$, may be
interpreted as due to a departure from Einstein gravity or, alternatively, due to a dark
energy component with anisotropic shear. 3) We employ a subset of Einstein's equations to
evolve linear perturbations under the presumed non-GR theory, to be defined shortly. Hence, we
are making the implicit assumption that a portion of the new field equations are equivalent to
GR with an additional shear. 

\begin{figure}[b] 
\includegraphics[scale=0.5]{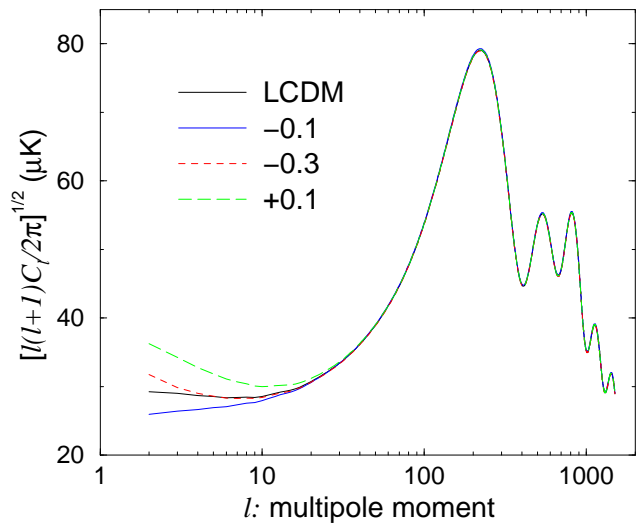}
\caption{The CMB anisotropy spectra under recipe R1 are shown for different values of
$\varpi_0$. R3 produces identical spectra. The backgrounds for all models are identical:
$\Omega_m = 0.35$, $h=0.65$ and dark energy with $w=-1$.}
\label{fig:R1cmb}
\end{figure}

We pause to emphasize the distinction between dark energy with shear and departures from GR.
Although the slip between the gravitational potentials can arise from either source, non-GR
theories of gravity are not generally equivalent to GR and dark energy with shear. Our
implementation of $\varpi$ represents a only portion of the non-standard behavior resulting
from different gravitational theories. Since observations already tightly constrain gross
departures from GR, we do not expect to be missing much with our parametrization. To keep the
phenomenology simple, we restrict attention to $w=-1$. (As well, dark energy or modified
gravity with $w\neq -1$ would surely be detected through classical tests of cosmology before
the effects of shear are manifest.)

Different theories of gravitation predict different relations between stress-energy and
spacetime curvature. Although we presume a $\Lambda$-like background evolution, the equations
guiding cosmological perturbations must differ. In the absence of a specific theory of
gravity, we intend to use the following recipe (R1):  3.i) Evolve radiation and matter fluid
sources using conservation of the individual stress-energy tensors, as in the standard case;
3.ii) Evolve $\varpi$ according to (\ref{eqn:varpi}); 3.iii) Enforce the slip between $\phi$
and $\psi$ by evolving $\alpha$ according to (\ref{eqn:alphadot}); 3.iv) Evolve the time-space
perturbed Einstein equation $k^2 \dot\eta = 4 \pi G a^2 (\rho+ p) \theta$ where $\theta$ is
due to the radiation and matter sources only, not dark energy.

Recipe R1 describes a theory of gravity in which the time-space equation is unchanged, but the
gravitational slip between $\phi$ and $\psi$ evolves as given by $\varpi$. Alternatively, this
describes a dark energy with shear perturbations but no fluctuations of the momentum density.
Implicitly, there must be energy density and pressure perturbations of the dark energy. (Note
that CMBfast \cite{Seljak:1996is} enforces energy conservation by applying the time-time
perturbed Einstein equation, $ \dot h = 2(k^2 \eta +4 \pi G a^2 \delta\rho)/{\cal H}$, at each
step. Instead, we use the definition of $\alpha$ to get $\dot h$.)

We have also considered alternate recipes. For recipe R2, the time-time perturbed Einstein
equation is evolved in place of the time-space equation. This corresponds to a theory of
gravity in which the time-time equation is unchanged; alternatively, the dark energy has no
energy density fluctuations, but does have momentum density and pressure perturbations. The
derivative $\dot\eta$ is obtained from the definition of $\alpha$. Finally, for recipe R3, we
drop the evolution of $\alpha$ and let the synchronous gauge variables $h$ and $\eta$ evolve
as in the standard case. However, for matters of calculating the CMB anisotropies we convert
to the conformal-Newtonian gauge with $\phi = \eta - {\cal H}\alpha$ but then artificially set
$\psi = (1 + \varpi)\phi$. 

\begin{figure}[t]
\includegraphics[scale=0.5]{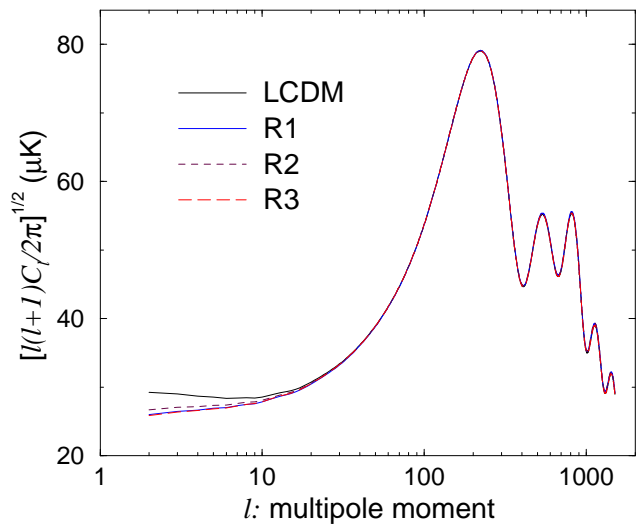}
\caption{The CMB anisotropy spectra under different recipes are shown for $\varpi_0=-0.1$.}
\label{fig:R123cmb}
\end{figure}

We have modified a version of CMBfast \cite{Seljak:1996is} to evolve perturbations according
to these schemes. The resulting CMB anisotropy spectra are shown in the following figures. All
the models are spatially flat with $\Omega_m = 0.35$, $h=0.65$, and normalized to have the
same small angular-scale power spectra as the corresponding $\Lambda$CDM model. In
Figure~\ref{fig:R1cmb} we show the anisotropy power spectrum for different values of
$\varpi_0$ under recipe R1. The alternative model, R3, produces identical spectra.
Empirically, we see that for small, negative values of $\varpi_0$ close to $-0.1$ there is a
cancellation between the ISW and SW effects which leads to a slight reduction in the amplitude
of the lowest multipole moments. For these cosmological parameters, the minimum value of the
quadrupole is achieved with $\varpi_0 \approx -0.15$. For large values of $|\varpi_0 + 0.15|$,
the reduction turns into an enhancement. In Figure~\ref{fig:R123cmb} we show the anisotropy
power spectrum with $\varpi_0 = -0.1$ for different recipes. All three recipes produce similar
results. 

The growth of structure is also modified by the gravitational slip. According to recipe R1,  
the cold dark matter density contrast $\delta_c$ evolves as
\begin{equation}
\ddot\delta_c + {\cal H}\dot\delta_c = k^2 (1 + \varpi)({\cal H}\alpha - \eta).
\end{equation}
where we have used $\dot\delta_c = -\dot h/2$, equation (\ref{eqn:alphadot}), and  we have
assumed that the shear due to radiation and neutrinos is negligible, as occurs beginning late 
in the matter-dominated era. Recipe R1 maintains the validity of the time-space perturbed
Einstein equation, whereas the time-time equation is implicitly modified. If the modifications
of the time-time equation are absorbed into a dark energy density perturbation, whereby  
$k^2\eta - {\cal H}\dot h/2 = -4 \pi G a^2(\delta\rho + \delta\rho_{DE})$, then the CDM
density contrast evolution equation becomes
\begin{equation}
\ddot \delta_c + {\cal H}\dot\delta_c = 4 \pi G a^2 (\delta\rho + \delta\rho_{DE})(1+\varpi).
\end{equation}
The effective dark energy density perturbation on the right-hand side is familiar from studies
of quintessence dark energy. The novel feature of this equation is the $1+\varpi$ factor which
can lead to faster (slower) growth for $\varpi > 0\, (<0)$. (Also see 
Refs.~\cite{Uzan:2000mz,Linder:2005in,Nesseris:2006er}.)

The modification of standard cosmological perturbation theory to incorporate the effects of
gravitational slip is not unique, as described above. There may be recipes other than R1-3 to
include $\varpi$ due to a departure from GR. Any particular non-GR theory makes a specific
prediction for the behavior of $\varpi$ as well as other effects. In the appendix, we study
the behavior of a dark energy component with $w=-1$ which also generates a gravitational slip.
However, as we have shown,   the differences in the three recipes are small when $\varpi_0$ is
small. Thus, despite the degeneracy, we find a practical advantage in this simple modeling of
departures from GR.

\section{CMB Anisotropy Spectrum}

\begin{figure}[b]
\includegraphics[scale=0.45]{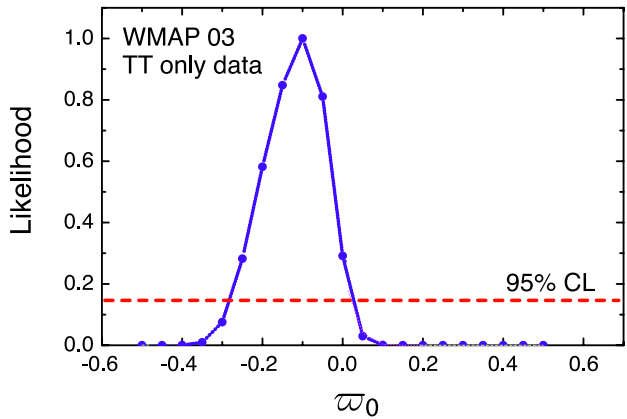}
\caption{Likelihood distribution for $\varpi_0$ due to WMAP 3-year CMB temperature 
measurements for a flat, ``concordance,'' $\Lambda$CDM model with $\Omega_{\Lambda}=0.76$,
$\Omega_b=0.02$, $h=0.73$, adiabatic and scalar inflationary pertubations with a spectral
index of $n_s=0.958$, and an optical depth of $\tau=0.08$. The $95 \%$ CL range is $-0.28 <
\varpi_0 <0.05$ with a peak at a slightly negative value of $\varpi_0$ owing to the decrease in
large-scale power.}
\label{fig:likelihood}
\end{figure}

Let us now investigate to what extent $\varpi_0$ can be constrained from current CMB
observations. As shown in the previous section, a change in $\varpi_0$ will mostly affect the
large angular scale CMB anisotropies through the ISW effect, leaving the acoustic peak
structure unchanged. For a first study, we restrict the analysis to a flat, ``concordance,''
$\Lambda$CDM model with  $\Omega_{\Lambda}=0.76$, $\Omega_b=0.02$, $h=0.73$. We also make the
choice of adiabatic and scalar inflationary perturbations with a spectral index of $n_s=0.958$
and we fix the optical depth to $\tau=0.08$ as suggested by Wilkinson Microwave Anisotropy
Probe (WMAP) 3-year anisotropy and polarization measurements. We then let $\varpi_0$ vary and
compare the CMB power spectra with the WMAP three year temperature anisotropy
data~\cite{Spergel:2006hy,Hinshaw:2006ia,Page:2006hz,Jarosik:2006ib} (WMAP3). The likelihood
is determined using the October 2006 version of the WMAP likelihood code available at the
\texttt{LAMBDA} website \cite{lambda}. In Figure~\ref{fig:likelihood} we plot the likelihood
for $\varpi_0$ (under recipe R1) from WMAP. As we can see, the likelihood peaks at  $\varpi_0
\sim -0.1$ although $\varpi_0=0$ is consistent with the data: the $95 \%$ CL range is $-0.28 <
\varpi_0 <0.05$. The preference for negative values of $\varpi_0$ is a direct consequence of
the low measured CMB quadrupole. Hovewer, as shown in the previous section, very large
negative values such that $|\varpi_0|\gtrsim 0.5$ enhance the quadrupole, making the model
dramatically incompatible with the data. 

It is also interesting to investigate possible variations of $\varpi_0$ by relaxing the
assumption of $w=-1$ dark energy component. Let us indeed substitute the cosmological constant
with a perturbed fluid with constant equation of state $p=w\rho$ and sound speed $c_s^2$. An
equation of state $w \neq -1$ and perturbations in the fluid would also change the large
angular scale anisotropy. As it is well known (see e.g. ~\cite{Bond:1997wr,Huey:1998se}) a
geometrical degeneracy makes virtually impossible any determination of $w$ from the position
of the acoustic peaks in the CMB anisotropy spectrum. However, combined measurements constrain
the equation of state parameter to a conservative range $-1.2<w<-0.8$ (see e.g.
\cite{Spergel:2006hy}). In Figure~\ref{fig:likelihood} we show the likelihood for $\varpi_0$
when a dark energy fluid with equation of state $w=-0.8$ is considered. The other parameters
have been chosen to reproduce the acoustic peak structure of the concordance model through the
geometrical degeneracy relation. We therefore fix $\Omega_m=0.298$, $h=0.655$ and
$\Omega_b=0.0495$. We consider two possibilities for the sound speed parameter $c_s^2=1,\,0$. 
As we can see, the likelihood for $\varpi_0$ is only moderately affected: we obtain $-0.38
<\varpi_0 <0.05$ for $c_s^2=1$ and $-0.35 <\varpi_0 <0.09$ for $c_s^2=0$ at $95 \%$ CL. We can
therefore claim that current observations provide a conservative bound of $-0.4 <\varpi_0
<0.1$ at $95 \%$ CL. This constraint is stronger than the PPN, lensing, and x-ray constraints
on $\varpi_0$, based on our model (\ref{eqn:varpi}), because $\rho_{\rm DM} \ll \rho_{\rm M}$
on megaparsec scales and below.

\begin{figure}[h]
\includegraphics[scale=0.45]{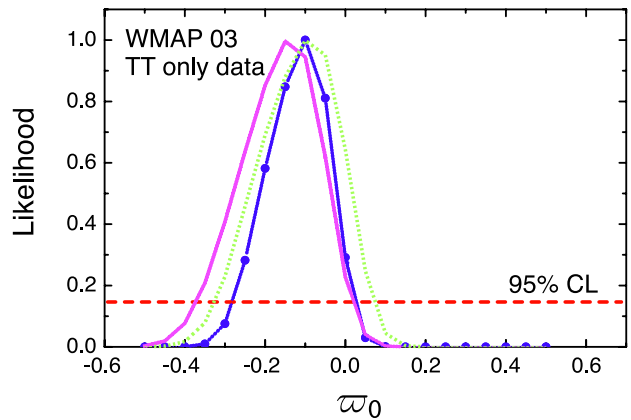}
\caption{Likelihood distributions for $\varpi_0$ due to WMAP 3-year CMB temperature 
measurements within the framework of different dark energy models.  The curve with dots is for
$\Lambda$CDM, as in the previous figure. The solid and dotted curves are for a dark energy
model with $w=-0.8$, with parameters chosen to reproduce the acoustic peak structure of the
$\Lambda$CDM model, but with dark energy perturbation sound speed $c_s^2=1,\,0$. The sound
speed does not have a strong effect on the likelihood range. Allowing for a range in equation
of state $-1.2 < w < -0.8$, then we obtain a conservative bound  $-0.4 <\varpi_0 <0.1$ at $95
\%$ CL.}
\label{fig:likelihood2}
\end{figure}

\section{Expectations for ISW and Weak Lensing Measurements}

Post-GR gravitational slip would also leave an imprint on the cross-correlation between the
CMB and large-scale structure. We define the two-point angular cross-correlation between the
temperature ISW anisotropy and the dark matter fluctuation as (see e.g. \cite{cora}):
\be
C^X(\theta)=\langle\Delta_{ISW}(\hat{\gamma_1})\delta_{LSS}(\hat{\gamma_2})\rangle,
\label{cross}
\ee
where the angular brackets denote the average over the ensemble and $\theta=\vert
\hat{\gamma_1}-\hat{\gamma_2}\vert$. For computational purposes it is convenient to decompose 
$C^X(\theta)$ into a Legendre series such that,
\be
C^X(\theta)=\sum_{l=2}^{\infty}\frac{2l+1}{4\pi}C_l^{X}P_l(\cos(\theta),
\label{cxpl}
\ee
where $P_l(\cos\theta)$ are the Legendre polynomials and $C_l^{X}$  is the cross-correlation
power spectrum given by \cite{Garriga:2003nm,Pogosian:2004wa}:
\be
C_l^X=4\pi\frac{9}{25}\int \frac{dk}{k}\Delta_{\mathcal{R}}^2 I^{ISW}_l(k)I^{LSS}_l(k),
\ee 
where $\Delta_{\mathcal{R}}^2$ is the primordial power spectrum. The integrand functions
$I^{ISW}_l(k)$ and $I^{LSS}_l(k)$ are defined respectively as:
\begin{eqnarray}
I^{ISW}_l(k)&=&-\int e^{-\kappa(z)} \frac{d((2+\varpi)\phi_k)}{dz} j_l[k r(z)] dz\\
I^{LSS}_l(k)&=&b\int  \Phi(z) \delta^k_m(z)j_l[k r(z)] dz,\label{growth}
\end{eqnarray}
where $\phi_k$ and $\delta_m^k$ are the Fourier components of the gravitational potential and
matter perturbation, respectively; $\Phi$ is the galaxy survey selection function; $j_l[k
r(z)]$ are the spherical Bessel functions; $r(z)$ is the comoving distance at redshift $z$ and
$\kappa(\tau)=\int_\tau^{\tau_0}\dot{\kappa}(\tau) d\tau$ is the total optical depth at time
$\tau$.

\begin{figure}[hb]
\includegraphics[scale=0.45]{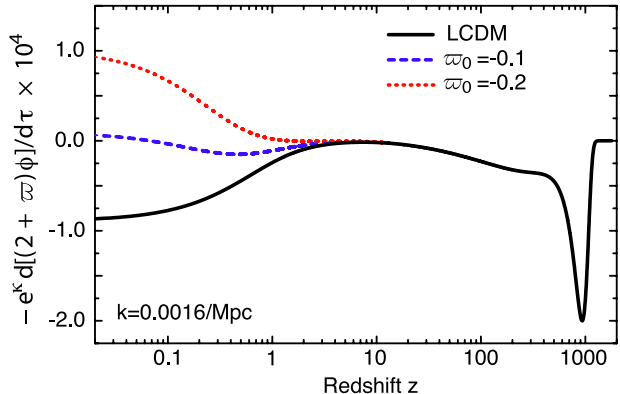}
\caption{Contributions to the integrated Sachs-Wolfe effect as a function of the redshift $z$
for different values of $\varpi_0$.}
\label{fig:iswint}
\end{figure}

A change in $\varpi$ could therefore change not only the value of $I^{ISW}_l$ but also its
sign respect to $I^{LSS}_l$. In Figure~\ref{fig:iswint}, we plot the values of the integrand  
$e^{-\kappa} d((2+\varpi)\phi_k)/d\tau$  at a scale of  $k\sim0.0016~{\rm Mpc}^{-1}$ as a
function of different values of $\varpi_0$. As we can see, while the early ISW at $z\sim 1000$
is left unchanged, a value of $\varpi_0\sim -0.2$ produces a change in the sign of the late
ISW contribution. This could indeed produce a negative cross-correlation with galaxy surveys.

\begin{figure}[ht]
\includegraphics[scale=0.45]{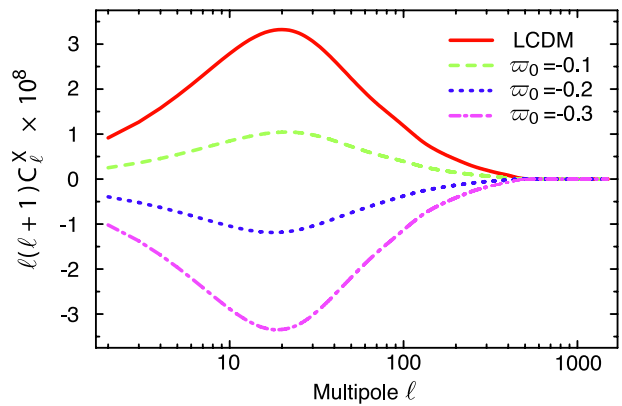}
\caption{Cross-correlation angular power spectrum between CMB temperature (ISW) and galaxy
distribution as a function of $\varpi_0$. A value of $\varpi_0<-0.2$ changes the sign of the
ISW, resulting in a negative cross-correlation, in conflict with observations.}
\label{fig5}
\end{figure}

Direct measurements of the cross-power spectrum $C_l^X$  are more robust for likelihood
parameter estimation since these data would be less correlated than measurements of
$C^X(\theta)$.  We therefore compute the cross-power spectrum $C_l^X$ for different values of
$\varpi_0$ assuming a galaxy survey with a selection function as
\begin{equation}
\Phi(z)\sim z^2 \exp{[-(z/\bar{z})^{1.5}]}
\end{equation} 
where $\bar{z}$, the median redshift of the survey, is  $\bar{z}=0.25$. In the past years, the
WMAP temperature anisotropy maps have been cross-correlated with several surveys of Large
Scale Structure (LSS) distributions and a positive correlation signal has been detected
\cite{Fosalba:2003ge,Boughn:2003yz,Scranton:2003in,Nolta:2003uy,Afshordi:2003xu,giannantonio06}.
As we can see from Figure~\ref{fig5}, values of  $\varpi_0 < -0.2$ would result in
anti-correlation and a negative angular spectrum which is in disagreement with current
observations at the $\sim 2 \sigma$ level.

\begin{figure}[ht]
\includegraphics[scale=0.4,angle=-90]{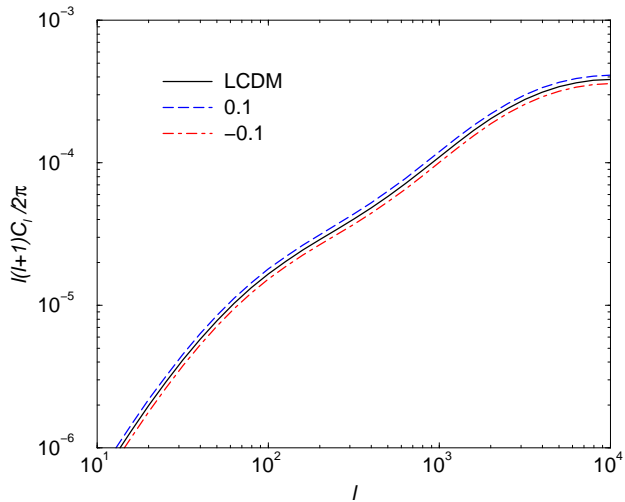}
\caption{Weak lensing convergence power spectrum for LCDM and for several values of $\varpi_0$
assuming that all sources are at $z_s=1$.}
\label{fig6}
\end{figure}

In addition to the ISW effect, weak lensing measurements of the large-scale structure capture
modifications imposed by $\varpi$. Using  equation~(\ref{eqn:lens}), we can write the angular
power spectrum of lensing convergence $\kappa$ as
\begin{equation}
C_l^{\kappa}= \frac{2}{\pi} \int k^2 dk\, \left[I^{\kappa}(k)\right]^2 P_{\phi \phi}(k) \, ,
\end{equation}
where
\begin{equation}
I^{\kappa}(k) = \int d\chi g(\chi) (2+\varpi)j_l(k\chi) \, .
\end{equation}
In Fig.~\ref{fig6},  we show the lensing convergence for our standard LCDM model and a model
in which $\varpi_0=-0.1$ and $\varpi_0=0.1$ in Eq.~(11)  assuming that all sources are at a
fixed redshift of $z_s=1$. As shown, there is an overall change in the amplitude of lensing
convergence fluctuations such that when $\varpi_0=-0.1$ (0.1)  the fluctuation power specrum
is lower (higher) by about 8\%. To see the modification associated with $\varpi$, the overall
normalization of the convergence power spectrum, such as through $\sigma_8$, must be known to
an accuracy better than 8\%. 

Detection of modifications from $\varpi$ with lensing statistics alone is impossible in the
same manner that lensing measurements of individual foreground objects cannot be used to
separate $\phi$ from $\psi$ as the measurements are sensitive to the combination only. To
constrain our post-GR parameter, as in Ref.~\cite{Bolton:2006yz}, one must combine lensing
measurements with an independent measurement which is sensitive to a different combination of
the two parameters or to just one of the parameters. In the case of galaxy lensing, the
comparison is made to dynamical mass estimates which are sensitive to $\phi$. In this context,
in the case of large-scale structure weak lensing, the weak lensing statistics must be
compared with an additional probe of large-scale structure, such as the potentials probed by
the galaxy distribution. While we have not carried out such an analysis here due to the as-yet
large uncertainties in weak lensing measurements, in the future it may be possible to further
constrain $\varpi_0$ through a combination of lensing measurements and galaxy clustering power
spectrum.

\section{Conclusions} 

We have introduced a new cosmological variable to characterize the degree of departure from
Einstein's General Relativity. The new parameter, $\varpi$, is the cosmological analog of the
parametrized post-Newtonian variable $\gamma$, which measures the amount of spacetime
curvature per unit mass.  Our study was motivated by various discussions in the literature on
uses of large-scale structure measurements to establish departures from GR on scales beyond
the solar system (e.g., \cite{Bertschinger:2006aw}). In the cosmological context, the
parameter $\varpi$ measures the difference between the Newtonian and longitudinal potentials
in response to the same matter sources, as occurs in certain scalar-tensor and other modified
theories of gravity. Equivalently, $\varpi$ measures the scalar shear fluctuation in a dark
energy component. In the context of a ``vanilla" LCDM background cosmology, a non-zero
$\varpi$ signals a departure from GR or a fluctuating cosmological constant. 

We have parameterized the time-evolution of $\varpi$ through a simple phenomenological model
in  which $\varpi=\varpi_0(1+z)^{-3}\Omega_{\rm DE}/\Omega_{\rm M}$.  The limit due to the
observed cosmic microwave background temperature anisotropies from the WMAP3 data
\cite{Spergel:2006hy,Hinshaw:2006ia,Page:2006hz,Jarosik:2006ib} is $-0.4 < \varpi_0 < 0.1$ at
the 95\% confidence level. The detection of a positive ISW effect
\cite{Fosalba:2003ge,Boughn:2003yz,Scranton:2003in,Nolta:2003uy,Afshordi:2003xu,giannantonio06}
further limits $\varpi_0 > -0.2$. Within the context of our model for $\varpi(z)$, these
constraints are tighter than the solar system time-delay and galaxy-scale gravitational
lensing tests of GR by an order of magnitude.  

We have considered possibilities to improve these constraints and have discussed the potential
use of both ISW and weak lensing measurements of the large-scale structure. In both cases,
these measurements must be combined with additional probes of large-scale structure, such as
the galaxy power spectrum, to break degeneracies between $\varpi$ and parameters. While
existing data are limited given the large uncertainities involved, weak lensing measurements
combined with galaxy clustering may be an ideal probe of $\varpi$. Further evidence that
$|\varpi| \ll 1$, or a detection of $\varpi \neq 0$, will provide useful clues as to the
validity of GR and the nature of dark energy.

\acknowledgments
We thank Caltech for hospitality, where a portion of this work was completed. R.C. was
supported in part by NSF AST-0349213 at Dartmouth. A.M. wishes to thank Pierstefano Corasaniti
for useful discussions.

\appendix

\section{w=-1 Dark Energy Fluctuations}

It is an apparent contradiction that a cosmological constant can fluctuate. Einstein's
cosmological term is regarded as a universal constant, leading to a stress-energy
$T^{\mu}_{\nu} = -\Lambda g^\mu_\nu/8\pi G$. The fluctuations must vanish by definition,
$\delta T^{\mu}_{\nu} = 0$. To avoid confusion, we will keep the cosmological constant
pristine  -- anything otherwise would be a misnomer. However, if the dark energy is a constant
only after some suitable averaging or expectation value is obtained, then it is not clear that
its fluctuations must also vanish. An example which has been explored recently is the ghost
condensate \cite{Arkani-Hamed:2003uy}, consisting of an effective field theory which gives
rise to a cosmological fluid with equation-of-state $w=-1$, but also supports scalar
fluctuations and a novel dispersion relation. The full implications of such a theory are
beyond the scope of our investigation, but serves as a motivating proof-of-principle.

In the absence of a specific model, we describe the dark energy as an imperfect fluid with
equation-of-state $w=p/\rho$ for the homogeneous, isotropic background, but which also
supports anisotropic stress or shear, $\sigma$.  Examples of dark energy models with
anisotropic stress are given in
Refs.~\cite{Battye:2006mb,Koivisto:2005mm,Gordon:2004ez,Hu:1998kj}. The equations of motion
for the linearized perturbations are given by
\begin{eqnarray}
\dot\theta &=& -{\cal H}(1 - 3 w)\theta - \frac{\dot w}{1+w}\theta + \frac{\delta p /
\delta\rho}{1+w}k^2\delta - k^2\sigma  + k^2\psi \cr
\dot\delta &=& -(1+w)(\theta - 3 \dot\phi) - 3{\cal H}\left(\frac{\delta p}{\delta\rho} -
w\right)\delta 
\end{eqnarray}
which are merely equations (30a, b) of \cite{Ma:1995ey}. Note that we neglect vector and
tensor perturbations, which may play a role in dark energy models with shear. The pressure and
energy density perturbations are linked by $c_{\rm s}^2 \equiv \delta p / \delta\rho$. If we
consider that ${\cal V}= (1+w)\theta$ is the relevant variable to describe the momentum
density perturbation, and $\Pi = (1+w)\sigma$ describes the shear, then the equations become
\begin{eqnarray}
\dot{\cal V} &=& -{\cal H}(1 - 3 w){\cal V} + c_{\rm s}^2 k^2 \delta  - k^2 \Pi + (1+w) k^2 \psi 
\cr
\dot\delta &=& -{\cal V} + 3(1+w) \dot\phi - 3{\cal H}\left(c_{\rm s}^2 - w\right)\delta .
\end{eqnarray}
Next, we set $w=-1$ and make the reasonable assumption that the gravitational potentials
$\phi$ and $\psi$ remain finite but that ${\cal V},\,\delta \rho,\,\delta p,\,\Pi$ do not
automatically vanish. The equations become
\begin{eqnarray}
\dot{\cal V} &=& -4 {\cal H}{\cal V} + c_{\rm s}^2 k^2 \delta  - k^2 \Pi, \cr
\dot\delta &=& -{\cal V} - 3{\cal H}\left(c_{\rm s}^2 + 1\right)\delta 
\label{eqn:newperts}
\end{eqnarray}
There is no longer any source term representing the inhomogeneities of the surrounding medium;
the fluctuations only feel the background, cosmic expansion. Of course, these fluctuations
still feed in to the metric perturbation equations. Note that the set of equations
(\ref{eqn:newperts}) become trivial in the case of a scalar field, for which $\Pi=\delta={\cal
V}=0$ in the limit $w\to -1$.

The system of equations (\ref{eqn:newperts}) can be expressed as a single, second order
equation for $\delta$:
\begin{equation}
\ddot\delta + (7 + 3 c_{\rm s}^2){\cal H}\dot\delta + \left[(12 {\cal H}^2 + 3 {\cal H}') (1 + c_{\rm s}^2)
+ c_{\rm s}^2 k^2 \right]\delta = k^2 \Pi .
\end{equation}
In the absence of shear, the short-wavelength modes are stable provided $c_{\rm s}^2 \ge 0$. 
The long-wavelength modes decay as $a^{-3},\,a^{-3(1+c_{\rm s}^2)}$ in a radiation-dominated 
epoch, and as $a^{-7/2},\,a^{-3(1+c_{\rm s}^2)}$ under matter. Any such fluctuations
are of little interest since the decay is so rapid. 

A wider variety of behavior is possible with the addition of shear. The sound speed $c_{\rm
s}$ no longer serves as the relevant quantity guiding the propagation and stability of
perturbations. Now, $c_{\rm s}^2\delta \ge \Pi$ is required for stability, and the
high-frequency phase velocity is $v = (c_{\rm s}^2 - \Pi/\delta)^{1/2}$.  This corresponds 
to a time and scale-dependent cosmic post-GR parameter
\begin{equation}
\varpi = -12 \pi G a^2 \rho \Pi / k^2 \phi.
\end{equation}
The propserties of the shear, and therefore the cosmic post-GR parameters, are dictated by the
details of any particular model. The quartic dispersion relation for the excitations in the
ghost condensate can be modeled by
\begin{equation}
\Pi = -\frac{k^2}{M^2} \delta.
\end{equation}
where $M$ is the energy scale below which the effective field theory is valid. Stability of
modes within the horizon requires $c_{\rm s}^2\ge 0$, in which case the perturbations decay
rapidly. Next, a phenomenological evolution equation has been proposed by Hu \cite{Hu:1998kj}, whereby
\begin{equation}
\dot\Pi + 3 {\cal H}\Pi = \frac{8}{3}c_{\rm vis}^2 \theta  
\label{eqn:huspi}
\end{equation}
(when translated into our notation) in the conformal-Newtonian gauge. The viscosity sound
speed $c_{\rm vis}$ modulates the source of the shear, which otherwise decays. However, a
phenomenological model which reflects equation (\ref{eqn:localratio}), whereby the 
shear-effects turn on as the local dark-energy density overtakes the matter, can be achieved
if
\begin{equation}
\Pi \equiv \frac{1}{3}\varpi_0 \delta.
\label{eqn:pi2delta}
\end{equation}
In this case, $\varpi \approx \varpi_0 \delta\rho_{DE}/\delta\rho_{TOT}$ on scales within the
horizon, which resembles our model (\ref{eqn:varpi}). We can also make a connection with Hu's
formula by plugging (\ref{eqn:pi2delta}) into (\ref{eqn:newperts}), such that
\begin{equation}
\dot\Pi + 3 {\cal H}(1 + c_{\rm s}^2 - \frac{1}{3}\frac{d\ln \varpi_0}{d \ln a})\Pi 
= -\frac{1}{3}\varpi_0 {\cal V}.  
\end{equation}
This corresponds to a viscosity sound speed $c_{\rm vis}^2 = -\varpi_0 (1+w)/8$, but a damping
rate which is now variable. With $c_{\rm s}^2 < -1$ and a constant or growing parameter
$\varpi_0$, the anisotropic scalar shear can actually grow. It is not clear what would be an
appropriate choice of initial conditions for fluctuations in this dark energy component, since
adiabaticity requires $\delta = {\cal V}=0$. Initial conditions which are set by some measure
of the relative contribution to curvature perturbations, $\delta R = -8\pi G(\delta\rho - 3
\delta p)$, must be finely tuned in order to force a non-negligible gravitational slip at late
times. In this investigation, however, we have focused directly on the slip rather than
fluctuating dark energy models.




\begin{thebibliography}{99}


\bibitem{Peebles:2002gy}
  P.~J.~E.~Peebles and B.~Ratra,
  Rev.\ Mod.\ Phys.\  {\bf 75}, 559 (2003).

\bibitem{Padmanabhan:2002ji}
  T.~Padmanabhan,
  Phys.\ Rept.\  {\bf 380}, 235 (2003).
    
\bibitem{Sahni:1999gb}
  V.~Sahni and A.~A.~Starobinsky,
  Int.\ J.\ Mod.\ Phys.\  D {\bf 9}, 373 (2000).

\bibitem{Carroll:2003wy}
  S.~M.~Carroll, V.~Duvvuri, M.~Trodden and M.~S.~Turner,
  Phys.\ Rev.\  D {\bf 70}, 043528 (2004).
  
\bibitem{Carroll:2004de}
  S.~M.~Carroll, A.~De Felice, V.~Duvvuri, D.~A.~Easson, M.~Trodden and M.~S.~Turner,
  Phys.\ Rev.\  D {\bf 71}, 063513 (2005).
  
\bibitem{Ishak:2005zs}
  M.~Ishak, A.~Upadhye and D.~N.~Spergel,
  Phys.\ Rev.\  D {\bf 74}, 043513 (2006).

\bibitem{Lue:2003ky}
  A.~Lue, R.~Scoccimarro and G.~Starkman,
  Phys.\ Rev.\  D {\bf 69}, 044005 (2004).

\bibitem{Zhang:2005vt}
  P.~Zhang,
  Phys.\ Rev.\  D {\bf 73}, 123504 (2006).

\bibitem{Knox:2006fh}
  L.~Knox, Y.~S.~Song and J.~A.~Tyson,
  Phys.\ Rev.\  D {\bf 74}, 023512 (2006). 

\bibitem{Uzan:2006mf}
  J.~P.~Uzan,
  arXiv:astro-ph/0605313.

\bibitem{Huterer:2006mv}
  D.~Huterer and E.~V.~Linder,
  Phys.\ Rev.\  D {\bf 75}, 023519 (2007).    

\bibitem{Tegmark:2001zc}
  M.~Tegmark,
  Phys.\ Rev.\  D {\bf 66}, 103507 (2002).

\bibitem{Sealfon:2004gz}
  C.~Sealfon, L.~Verde and R.~Jimenez,
  Phys.\ Rev.\  D {\bf 71}, 083004 (2005).

\bibitem{Shirata:2005yr}
  A.~Shirata, T.~Shiromizu, N.~Yoshida and Y.~Suto,
  Phys.\ Rev.\  D {\bf 71}, 064030 (2005).
    
\bibitem{Jaekel:2005qe}
  M.~T.~Jaekel and S.~Reynaud,
  Class.\ Quant.\ Grav.\  {\bf 22}, 2135 (2005).
  
\bibitem{Jaekel:2005qz}
  M.~T.~Jaekel and S.~Reynaud,
  Class.\ Quant.\ Grav.\  {\bf 23}, 777 (2006). 
   
\bibitem{WillBook}
  C.~M.~Will,
  ``Theory and Experiment in Gravitational Physics,''
  (Cambridge: Cambridge University Press), (1993).
 
\bibitem{Will:2001mx}
  C.~M.~Will,
  Living Rev.\ Rel.\  {\bf 4}, 4 (2001).
   
\bibitem{Williams:2004}
  J.~G.~Williams, S.~G.~Turyshev, and D.~H.~Boggs,
  Phys. Rev. Lett. {\bf 93}, 261101 (2004). 

\bibitem{Bertotti:2003}
  B.~Bertotti, L.~Iess and P.~Tortora,
  Nature {\bf 425}, 374 (2003). 

\bibitem{Bekenstein:1993fs}
  J.~D.~Bekenstein and R.~H.~Sanders,
  Astrophys.\ J.\  {\bf 429}, 480 (1994).

\bibitem{White:2001kt}
  M.~J.~White and C.~S.~Kochanek,
  Astrophys.\ J.\  {\bf 560}, 539 (2001).

\bibitem{Bolton:2006yz}
  A.~S.~Bolton, S.~Rappaport and S.~Burles,
  Phys.\ Rev.\  D {\bf 74}, 061501 (2006).

\bibitem{Ma:1995ey}
  C.~P.~A.~Ma and E.~Bertschinger,
  Astrophys.\ J.\  {\bf 455}, 7 (1995). 

\bibitem{Wu98}
  X.~P.~Wu, T.~Chiueh, L.~Z.~Fang and Y.~J.~Xue,
  Mon.\ Not.\ Roy.\ Astron.\ Soc.\ {\bf 301}, 861 (1998).
  
\bibitem{Acquaviva:2004fv}
  V.~Acquaviva, C.~Baccigalupi and F.~Perrotta,
  Phys.\ Rev.\ D {\bf 70}, 023515 (2004).

\bibitem{Bertschinger:2006aw}
  E.~Bertschinger,
  Astrophys.\ J.\  {\bf 648}, 797 (2006).

\bibitem{Chen:1999qh}
  X.~l.~Chen and M.~Kamionkowski,
  Phys.\ Rev.\  D {\bf 60}, 104036 (1999).
  
\bibitem{Esposito-Farese:2000ij}
  G.~Esposito-Farese and D.~Polarski,
  Phys.\ Rev.\  D {\bf 63}, 063504 (2001).

\bibitem{Riazuelo:2001mg}
  A.~Riazuelo and J.~P.~Uzan,
  Phys.\ Rev.\  D {\bf 66}, 023525 (2002)  
 
\bibitem{Nagata:2002tm}
  R.~Nagata, T.~Chiba and N.~Sugiyama,
  Phys.\ Rev.\  D {\bf 66}, 103510 (2002).
  
\bibitem{Nagata:2003qn}
  R.~Nagata, T.~Chiba and N.~Sugiyama,
  Phys.\ Rev.\  D {\bf 69}, 083512 (2004).
  
\bibitem{Schimd:2004nq}
  C.~Schimd, J.~P.~Uzan and A.~Riazuelo,
  Phys.\ Rev.\ D {\bf 71}, 083512 (2005).

\bibitem{Koivisto:2005mm}
  T.~Koivisto and D.~F.~Mota,
  Phys.\ Rev.\  D {\bf 73}, 083502 (2006)

\bibitem{Battye:2006mb}
  R.~A.~Battye and A.~Moss,
  Phys.\ Rev.\  D {\bf 74}, 041301 (2006).

\bibitem{Capozziello:2006fa}
  S.~Capozziello, A.~Stabile and A.~Troisi,
  Mod.\ Phys.\ Lett.\  A {\bf 21}, 2291 (2006). 
  
\bibitem{Amendola:2005cr}
  L.~Amendola, C.~Charmousis and S.~C.~Davis,
  JCAP {\bf 0612}, 020 (2006).
        
\bibitem{Dvali:2000rv}
  G.~R.~Dvali, G.~Gabadadze and M.~Porrati,
  Phys.\ Lett.\  B {\bf 484}, 112 (2000). 

\bibitem{Lue:2005ya}
  A.~Lue,
  Phys.\ Rept.\  {\bf 423}, 1 (2006).
 
\bibitem{Koyama:2005kd}
  K.~Koyama and R.~Maartens,
  JCAP {\bf 0601}, 016 (2006).
   
\bibitem{Lue:2004rj}
  A.~Lue, R.~Scoccimarro and G.~D.~Starkman,
  Phys.\ Rev.\  D {\bf 69}, 124015 (2004).
   
\bibitem{Seljak:1996is}
  U.~Seljak and M.~Zaldarriaga,
  Astrophys.\ J.\  {\bf 469}, 437 (1996).

\bibitem{Uzan:2000mz}
  J.~P.~Uzan and F.~Bernardeau,
  Phys.\ Rev.\  D {\bf 64}, 083004 (2001).
     
\bibitem{Linder:2005in}
  E.~V.~Linder,
  Phys.\ Rev.\  D {\bf 72}, 043529 (2005).   
   
\bibitem{Nesseris:2006er}
  S.~Nesseris and L.~Perivolaropoulos,
  JCAP {\bf 0701}, 018 (2007).   
   
\bibitem{Spergel:2006hy}
  D.~N.~Spergel {\it et al.},
  arXiv:astro-ph/0603449.

\bibitem{Hinshaw:2006ia}
  G.~Hinshaw {\it et al.},
  arXiv:astro-ph/0603451.

\bibitem{Page:2006hz}
  L.~Page {\it et al.},
  arXiv:astro-ph/0603450.

\bibitem{Jarosik:2006ib}
  N.~Jarosik {\it et al.},
  arXiv:astro-ph/0603452.

\bibitem{lambda}
  \texttt{http://lambda.gsfc.nasa.gov}

\bibitem{Bond:1997wr}
  J.~R.~Bond, G.~Efstathiou and M.~Tegmark,
  Mon.\ Not.\ Roy.\ Astron.\ Soc.\  {\bf 291}, L33 (1997).
   
\bibitem{Huey:1998se}
  G.~Huey, L.~M.~Wang, R.~Dave, R.~R.~Caldwell and P.~J.~Steinhardt,
  Phys.\ Rev.\  D {\bf 59}, 063005 (1999)
  
\bibitem{cora}
  P.~S.~Corasaniti, T.~Giannantonio and A.~Melchiorri,
  Phys.\ Rev.\  D {\bf 71}, 123521 (2005). 

\bibitem{Garriga:2003nm}
  J.~Garriga, L.~Pogosian and T.~Vachaspati,
  Phys.\ Rev.\  D {\bf 69}, 063511 (2004).
  
\bibitem{Pogosian:2004wa}
  L.~Pogosian,
  JCAP {\bf 0504}, 015 (2005).
  
\bibitem{Fosalba:2003ge}
  P.~Fosalba, E.~Gaztanaga and F.~Castander,
  Astrophys.\ J.\  {\bf 597}, L89 (2003).

\bibitem{Boughn:2003yz}
  S.~Boughn and R.~Crittenden,
  Nature {\bf 427}, 45 (2004).
  
\bibitem{Scranton:2003in}
  R.~Scranton {\it et al.}  [SDSS Collaboration],
  arXiv:astro-ph/0307335.
    
\bibitem{Nolta:2003uy}
  M.~R.~Nolta {\it et al.}  [WMAP Collaboration],
  Astrophys.\ J.\  {\bf 608}, 10 (2004).
  
\bibitem{Afshordi:2003xu}
  N.~Afshordi, Y.~S.~Loh and M.~A.~Strauss,
  Phys.\ Rev.\  D {\bf 69}, 083524 (2004).

\bibitem{giannantonio06}
  T.~Giannantonio {\it et al.},
  Phys.\ Rev.\  D {\bf 74}, 063520 (2006). 

\bibitem{Arkani-Hamed:2003uy}
  N.~Arkani-Hamed, H.~C.~Cheng, M.~A.~Luty and S.~Mukohyama,
  JHEP {\bf 0405}, 074 (2004),
     
\bibitem{Gordon:2004ez}
  C.~Gordon and W.~Hu,
  Phys.\ Rev.\ D {\bf 70}, 083003 (2004).
  
\bibitem{Hu:1998kj}
  W.~Hu,
  Astrophys.\ J.\  {\bf 506}, 485 (1998).
       
     
\end{thebibliography}
\end{document}